\documentclass[aip,jcp,reprint]{revtex4-1}

\usepackage{graphicx}
\usepackage{amsmath,amssymb,amstext,graphicx,bm,booktabs,hyperref,siunitx,upgreek}

\newcommand{\natcs}{\ensuremath{\mathrm{N@C}_{60}}}
\newcommand{\fnatcs}{\ensuremath{^{15}\mathrm{N@C}_{60}}}
\newcommand{\cs}{$\mathrm{C}_{60}$}

\begin{document}

\title{High pressure electron spin resonance of the endohedral fullerene $^{\mathbf {15}}$N@C$_\mathbf{60}$}

\author{R.T.~Harding}
\affiliation{Department of Materials, University of Oxford, Parks Road, Oxford OX1 3PH, United Kingdom}

\author{A.~Folli}
\affiliation{School of Chemistry, Cardiff University, Cardiff CF10 3AT, United Kingdom}

\author{J.~Zhou}
\affiliation{Department of Materials, University of Oxford, Parks Road, Oxford OX1 3PH, United Kingdom}

\author{G.~A.~D.~Briggs}
\affiliation{Department of Materials, University of Oxford, Parks Road, Oxford OX1 3PH, United Kingdom}

\author{K.~Porfyrakis}
\affiliation{Department of Materials, University of Oxford, Parks Road, Oxford OX1 3PH, United Kingdom}

\author{E.A.~Laird}
\affiliation{Department of Materials, University of Oxford, Parks Road, Oxford OX1 3PH, United Kingdom}

\begin{abstract}
We measure the electron spin resonance spectrum of the endohedral fullerene molecule \fnatcs~at pressures ranging from atmospheric pressure to 0.25\,GPa, and find that the hyperfine coupling increases linearly with pressure.
We present a model based on van der Waals interactions, which accounts for this increase via compression of the fullerene cage and consequent admixture of orbitals with a larger hyperfine coupling.
Combining this model with theoretical estimates of the bulk modulus, we predict the pressure shift and compare it to our experimental results, finding fair agreement given the spread in estimates of the bulk modulus.
The spin resonance linewidth is also found to depend on pressure.
This is explained by considering the pressure-dependent viscosity of the solvent, which modifies the effect of dipolar coupling between spins within fullerene clusters. 

\end{abstract}

\date{\today}

\maketitle

\section{Introduction}
In the central-field approximation, atomic nitrogen has no hyperfine coupling because the combined spin density of the three unpaired $p$ orbitals vanishes at the nucleus and is spherically symmetric outside it.\cite{Holloway1962}
In reality, interactions beyond this approximation lead to a non-zero hyperfine coupling for the free atom.\cite{Heald1954}
In the presence of other atoms or molecules, distortion of the electron orbitals further modifies the hyperfine coupling, which therefore offers an insight into the nature of the inter-atomic interactions.\citep{Adrian1962}

The endohedral fullerene \natcs~offers a nearly unique system in which the fullerene cage stabilises the encapsulated nitrogen such that it behaves like a free atom.\cite{AlmeidaMurphy1996}
However, the hyperfine coupling of \natcs~is enhanced by approximately 50\% relative to the free atomic value, which reflects the effect of confinement of the nitrogen orbitals by the cage. This increase has previously been attributed to nitrogen--cage interactions that mix in excited states with larger hyperfine couplings, although without proposing a microscopic model.\cite{AlmeidaMurphy1996}

Understanding the hyperfine coupling of this molecule is important due to its potential use as a molecular spin qubit\cite{PhysRevA.65.032322} or frequency reference in an atomic clock.\cite{briggs2012atomic}
In particular, \fnatcs~is suitable as a frequency reference due to its sharp resonances\cite{Morton2006} and the existence of a clock transition in its low-field spectrum.\cite{PhysRevLett.119.140801}
Since the frequency of this clock transition depends solely on the isotropic hyperfine coupling constant $A$, it is important to characterise the mechanisms that affect it.
For example, the hyperfine coupling is known to depend on temperature,\cite{Pietzak1998} which could reduce the long term stability of a fullerene clock against environmental temperature fluctuations.
 
Here, we measure the hyperfine coupling strength and the electron spin resonance (ESR) linewidth over a pressure range up to 0.25\,GPa.
The hyperfine coupling increases linearly with pressure, which we explain using a microscopic model of electron wavefunction distortion mediated by van der Waals interactions between the nitrogen atom and the cage.
Using this model and theoretical estimates of the bulk modulus of isolated fullerene molecules,\cite{Ruoff1991,Woo1992,Kaur2010,Peon-Escalante2014,AMER2009232} we then estimate the pressure shift and compare it with our experimental results.
The predicted shift is smaller than the observed shift, which may indicate contributions beyond the van der Waals interaction as well as uncertainty in the the theoretical bulk modulus.
To the best of our knowledge, the value deduced from our model is the first experimental estimate of bulk modulus for an individual \cs~molecule under conditions of hydrostatic pressure.

We find that the measured linewidth increases non-linearly with pressure.
This is explained by the pressure-dependent viscosity of the solvent\cite{VieiradosSantos1997} and its effect on rotational diffusion\cite{Bloembergen1948} of fullerene clusters.\cite{Guo2016,Dattani2015,Makhmanov2016}
At low pressure, the solvent is sufficiently non-viscous that dipole-dipole coupling between spins is averaged out by rotation of the cluster.
At high pressure, the solvent viscosity increases, leading to an increase in the linewidth towards the rigid lattice limit.\cite{Edelstein1964}

\section{Hyperfine coupling}
\subsection{Experiment}
A \fnatcs~sample was prepared by ion implantation,\cite{AlmeidaMurphy1996} dissolved in toluene, and purified by high-performance liquid chromatography to a purity (\natcs:\cs) of $\sim$0.6\,\%.\cite{Kanai2004}
The sample used for high-pressure measurements was then concentrated by bubbling nitrogen gas through it to evaporate some solvent.
This is because the small sample volume of the pressure cell leads to a low resonator filling factor, which reduces the signal-to-noise ratio (SNR) of the measurement.
This sample was then injected into an yttria-stabilised zirconia cell attached to a barocycler and loaded into a dielectric resonator (Bruker ER4123D).
Measurements were performed at the $X$ band (frequency $f\sim9.5$\,GHz) using a Bruker EMXmicro spectrometer.
All measurements were performed at room temperature, with sufficient time between increasing the pressure and performing the measurement to allow thermal equilibriation.

The energy levels of the \fnatcs~spin system are described by the Hamiltonian
\begin{equation}\label{Equation1}
\mathcal{H}=g_\mathrm{e} \mu_\mathrm{B} S_z B_0 - g_\mathrm{N} \mu_\mathrm{N} I_z B_0 + A\hat{\boldsymbol{S}}\cdot\hat{\boldsymbol{I}},
\end{equation}
where the first two terms parameterise the Zeeman interaction of the electron spin $S=3/2$ and nuclear spin $I=1/2$ with the static magnetic field $B_0$.
The electronic and nuclear $g$ factors are denoted by $g_e$ and $g_N$, respectively.
The final term describes the hyperfine interaction, with coupling strength $A$.
At high fields, where the Zeeman interaction dominates, the energy levels are labelled by the quantum numbers $m_S$ and $m_I$, which give the projection of the relevant spin onto the axis defined by the direction of $B_0$.
The ESR signals arise from magnetic dipole transitions between different $m_S$ levels, with selection rules $\Delta m_S = \pm 1$.

\begin{figure}
\includegraphics[width=\columnwidth]{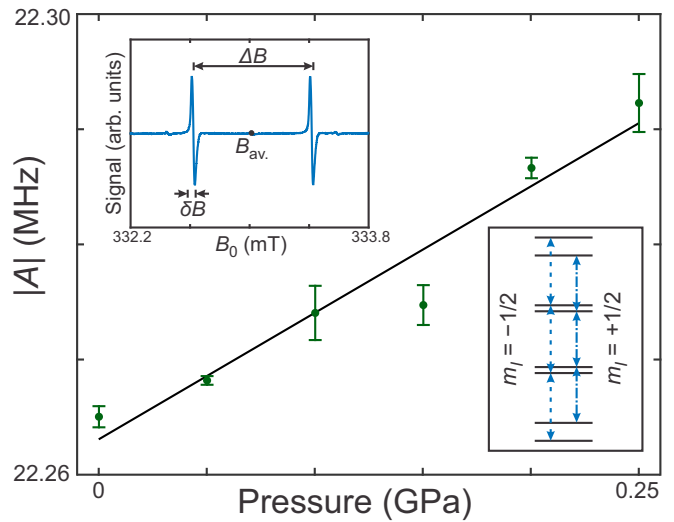}
\caption{Hyperfine coupling constant $A$ as function of pressure. The data (dark green circles with error bars) are fitted by a linear function (solid black line).  The data points are the mean value of $A$ at each pressure, and the error bars are the standard deviation of this measurement. Upper inset: ESR signal as function of magnetic field at frequency $f\approx 9.334$\,GHz, demonstrating the hyperfine splitting $\Delta B$, peak-to-peak linewidth $\delta B$, and the centre field $B_\mathrm{av}$. Lower inset: schematic of high-field energy levels showing the two sets of transition frequencies corresponding to the two values of $m_I$.
\label{Figure1}
}
\end{figure}

The ESR spectrum, measured as a function of magnetic field, exhibits a pair of spin resonances corresponding to the two states of the $^{15}$N nuclear spin, as shown in the upper inset of Fig.~\ref{Figure1}.
The resonances are separated in the field domain by the hyperfine splitting $\Delta B$.
We extract $\Delta B$ by fitting the spectrum to identify the extrema of each resonance observed in the first-derivative spectra.
This splitting is converted to a hyperfine coupling using the relationship $A=g \mu_B \Delta B$, where the $g$ factor $g =h f/ \mu_\mathrm{B} B_\mathrm{av.}$.
Here, $B_\mathrm{av.}$ is the centre field of the spectrum and $f$ is the frequency of the applied microwave radiation.
This frequency depends weakly on pressure via the changing dielectric constant of toluene,\cite{Skinner1968,Mopsik1969} which alters the cavity frequency.

Data for the hyperfine coupling constant as a function of pressure are plotted in Fig.~\ref{Figure1}.
Fitting the data by a linear relationship, such that $A(P)=A_0+A_1P$, gives $A_0=22.263(2)$\,MHz and $A_1 = 1.1(1)\times10^{-4}$\,Hz\,Pa$^{-1}$.
The bracketed number gives the statistical error in the last digit of the quoted value.

\subsection{Theory}

We will now explain the pressure-dependent hyperfine coupling displayed in Fig.~\ref{Figure1}.
A similar effect has been measured previously for atomic nitrogen generated using a radio-frequency discharge plasma, where the hyperfine coupling is proportional to the buffer gas pressure.\cite{Holloway1962,Lambert1963}
In contrast to the pressure-dependent coupling observed for atoms with unpaired $s$ electrons, which can either be positive or negative, the magnitude of the nitrogen hyperfine coupling always increases with pressure.\cite{Adrian1962}

This observation has been explained by treating van der Waals interactions between the nitrogen atom and the buffer gas atoms as the dominant contribution to the increased hyperfine coupling.\cite{Adrian1962}
These interactions lead to spin-dependent excitation of the $s$ electrons, which causes spin polarization of the $s$ orbital at the nucleus and an increased hyperfine coupling.
Following an analogous model, we hypothesise that the increase in hyperfine coupling for the nitrogen endohedral fullerene is due to van der Waals interactions between the incarcerated nitrogen and the fullerene cage.
There should also be a pressure-dependent exchange coupling between the cage and the $p$ orbitals. However, because of spherical symmetry this does not mix $p$ and $s$ orbitals and is therefore expected to make a weaker contribution to the hyperfine coupling.\cite{Adrian1962} We therefore model the shift as arising entirely from van der Waals interactions.

To apply this model to the endohedral fullerene system, we consider the increase in hyperfine coupling $\Delta A=A-A_\mathrm{free}$, where $A_\mathrm{free}$ is the value for free atomic nitrogen.
A perturbative calculation of the van der Waals interaction between atomic nitrogen and a nearby particle predicts the relationship (Eq.~(15) of Ref.~\onlinecite{Adrian1962}):
\begin{equation}
\Delta A = \frac{k}{R^6},
\label{eq:delta_a}
\end{equation} where $R$ is the distance between the nitrogen and fullerene charge distributions.
Since the nitrogen atom is located at the centre of the cage,\cite{Pietzak1998} we set this distance equal to the fullerene radius.
The constant $k$ parameterises the strength of the interaction between the two charge clouds.
Performing an ab initio calculation of the value of $k$ is beyond the scope of this work; instead, we treat it as an experimentally determined parameter.

We test this model by using Eq.~(\ref{eq:delta_a}) to predict the pressure shift given the bulk modulus $B$.
By treating the cage as a sphere with volume $V$ and bulk modulus $B = -V \partial P / \partial V$, we obtain the expression
\begin{equation}
\left(\frac{\partial \Delta A}{\partial P}\right) = \frac{2 \Delta A}{B}.
\label{eq:bulk_modulus}
\end{equation}
For \fnatcs, we take~\cite{Lambert1963} $A_\mathrm{free}=-14.65$\,MHz and $A=-22.35$\,MHz and hence $\Delta A=7.70$\,MHz.
The bulk modulus of crystalline \cs~samples has been measured experimentally,\cite{Duclos1991,Lundin1994,Pintschovius1999,Levin2000} but the value for an isolated \cs~molecule under conditions of hydrostatic pressure is only known via simulations.\cite{Ruoff1991,Woo1992,Kaur2010,Peon-Escalante2014,AMER2009232}
The simulated values range from 300 to 1200\,GPa, with a grouping in the range 700 to 900\,GPa.\cite{Ruoff1991,Woo1992,Peon-Escalante2014}
Taking $B\sim800$\,GPa, Eq.~\ref{eq:bulk_modulus} predicts $\partial \Delta A / \partial P \sim 2 \times 10^{-5}$.
This value is approximately six times smaller than the value $\partial \Delta A / \partial P = 1.1(1) \times 10^{-4}$\,Hz\,Pa$^{-1}$ determined from the fit in Fig.~\ref{Figure1}.

The discrepancy could have at least two causes. It may indicate that exchange interaction between the nitrogen and the cage, not captured by our van der Waals model, in fact contributes significantly to the nitrogen hyperfine coupling. 
Alternatively, the discrepancy could indicate that previous predictions have overestimated the fullerene's bulk modulus.
To explain the measured pressure shift using Eq.~\ref{eq:bulk_modulus} would require a value $B=140 \pm 13$, which is outside the range of theoretical predictions. However, these predictions themselves cover a wide range, which may indicate a need for further modelling.
One source of uncertainty in the modelling is the effect of the solvent, which is expected to modify the molecule's bulk modulus.\cite{AMER2009232} In particular, the effect of toluene solvent has not yet been modelled.

\subsection{Solvent effects}

\begin{table}
\begin{tabular}{c c c c}
\noalign{\vspace{4pt}}
\hline\hline
\noalign{\smallskip}
Solvent & $\epsilon_r$ & $\mu$ (D)&  $|A|$ (MHz)\\
\hline \noalign{\smallskip}
Toluene & 2.39 & 0.375 & 22.247(2) \\
Carbon disulfide & 2.63 & 0 & 22.240(7)\\
Chlorobenzene & 5.69 & 1.69 & 22.253(5)\\ \noalign{\smallskip}
\hline\hline
\end{tabular}
\caption{Hyperfine coupling at room temperature and pressure in solvents with different dielectric constants $\epsilon_r$ and dipole moments $\mu$.
Values of the dielectric constant are given for atmospheric pressure and $T=293.2$\,K.\cite{crchandbook95} The bracketed number gives the one standard deviation error calculated by taking the sample standard deviation of six measurements.}
\label{Table1}
\end{table}

The most direct interpretation of the pressure dependence is that the nitrogen wavefunction is modified by compression of the cage.
However, for some dissolved radicals, the hyperfine coupling constant depends on the solvent due to interactions between the solvent molecules and the unpaired electrons of the radical.\cite{Owenius2001,Cook1983,Ludwig1964,Janzen1982,DAnna1970}
For \natcs, such an effect should be small, since the fullerene cage effectively isolates the atomic nitrogen from the environment.\cite{Dinse2000}
However, changes in the electronic properties of the solvent could plausibly alter the electronic properties of the fullerene cage; considering Eq.~\ref{eq:delta_a}, this would correspond to altering $k$.
Therefore, we must exclude the possibility that the pressure-dependent hyperfine coupling reflects changes in solvent properties, such as dielectric constant $\epsilon_r$ or molecular electric dipole moment $\mu$, that are known to vary with pressure.\cite{Skinner1968,Mopsik1969}

At room temperature and atmospheric pressure, toluene has dielectric constant $\epsilon_r=2.39$ and dipole moment $\mu=0.36$\,D.
At 0.25\,GPa, $\epsilon_r$ increases to $\sim$2.6, while $\mu$ is likely to vary by less than 2\%.
To measure the effect of changes in $\epsilon_r$ and $\mu$, we therefore compare the spectrum of \fnatcs~dissolved in toluene with those of \fnatcs~dissolved in chlorobenzene ($\epsilon_r=5.69$ and $\mu = 1.69$\,D) and carbon disulfide ($\epsilon_r=2.63$ and $\mu = 0$\,D).
The ESR spectra of \fnatcs~dissolved in these solvents were measured at room temperature and atmospheric pressure at the $X$ band using a Bruker ER4122-SHQE-W1 resonator and EMXmicro spectrometer.
Data from these experiments are presented in Table \ref{Table1}.

The data show that the effect of altering the dielectric properties of the solvent is negligible to within experimental error.
Moreover, the effect of using different solvents is less than the observed change in hyperfine coupling as a function of pressure, despite varying the dielectric constant and dipole moment by considerably more than the variation achieved by pressurising toluene.
The data were obtained using a different spectrometer to the data presented in Fig.~\ref{Figure1}, and the small difference in $A$ between comparable values presented in Fig.~\ref{Figure1} and Table~\ref{Table1} presumably reflects differences in magnet calibration.
Such a systematic error does not affect the validity of the comparison between different solvents shown here.
Therefore, we conclude that the increased hyperfine coupling at high pressures is not caused by pressure-dependent solvent properties, and that the cage compression model is correct.

\section{Linewidth}
\begin{figure}
\includegraphics[width=\columnwidth]{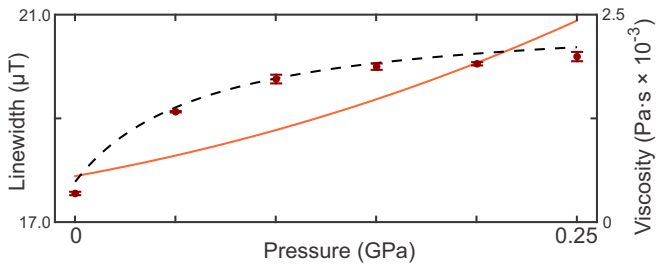}
\caption{Peak-to-peak linewidth and toluene viscosity as a function of pressure. The linewidth data (red circles with error bars) are fitted (black dashed line) using a model based on the rotation of fullerene clusters. The viscosity of toluene (solid orange line) over this pressure range is plotted from the measured equation of state given in reference \onlinecite{VieiradosSantos1997}.
\label{Figure2}
}
\end{figure}

In addition to the pressure-dependent hyperfine coupling, we also observe a pressure-dependent linewidth.
The field domain peak-to-peak linewidth $\delta B$, as shown in the upper inset of Fig.~\ref{Figure1}, was determined by measuring the splitting between the extrema of each resonance.
As shown in Fig.~\ref{Figure2}, the linewidth increases with pressure.

Pressure-dependent linewidths are typically governed by spin-exchange processes.\cite{Edelstein1964,Edelstein1964a,Gardiner1981,Mendt2015}
However, for \natcs~exchange interactions between the incarcerated nitrogen atoms are suppressed by the cage, which prevents overlap of the electronic wavefunctions.\cite{Dinse2000}
We therefore use a model that instead considers the dipole--dipole interactions between \fnatcs~spins in fullerene clusters.\cite{Guo2016,Dattani2015,Makhmanov2016}
Within each cluster, \fnatcs~spins interact via their magnetic moments, which leads to dipolar broadening.
At low pressures, the viscosity of the toluene solvent is low.\cite{Rowane2017,VieiradosSantos1997}
The fullerene cluster therefore rotates sufficiently rapidly to average out the dipolar coupling between the spins, leading to a narrow linewidth.
At high pressures, however, the viscosity increases, which reduces the rate of rotation. The linewidth then tends toward the rigid lattice limit imposed by dipole--dipole coupling between the spins in the cluster.\cite{Edelstein1964}

In this model, the dephasing time $T_2^*$ obeys the implicit equation
\begin{equation}
\left(T_2^*\right)^{-2} = (2/\pi) C^2 \tan^{-1} (2 \tau _c / T_2^*),
\label{eq:t2}
\end{equation}
where $\tau_c$ is the rotational correlation time and $C$ is the linewidth due to dipole-dipole broadening as $\tau_c \rightarrow \infty$, i.e.~the rigid lattice limit.\cite{Edelstein1964}
The rotational correlation time
$\tau_c = 4\pi \eta a^3 / 3 k_B T,$
where $\eta$ is the viscosity of the solvent, $a$ is the effective hydrodynamic radius of the rotating cluster, $k_B$ is the Boltzmann constant, and $T$ is the temperature.\cite{Bloembergen1948}
We fit the field domain linewidths $\delta B$ presented in Fig.~\ref{Figure2} as a function of solvent viscosity, which is known from the previously measured equation of state.\cite{VieiradosSantos1997} 
From the fit, we extract $C=590\pm100$\,kHz and $a = 15.5\pm 0.8$\,nm, which is comparable to the size of previously observed clusters.\cite{Makhmanov2016}

By considering the interaction energy of two spins, we determine the average spin--spin separation $r \approx \left(\mu_0 S(S+1) g_e^2 \mu_B^2/hC\right)^{1/3} = 16.1\pm 0.9$\,nm.

The value for $a$ is much greater than the radius of an individual \cs~molecule, while the value of $r$ is much less than the expected separation for unclustered fullerenes given the spin density. Both these facts are evidence for the clustering of fullerenes in our sample.
However, the observed spin--spin separation is larger than the separation expected if the cluster were formed of close packed fullerenes.
Such a cluster would have a spin--spin separation of approximately 5\,nm given the purity of the sample and the diameter of an individual fullerene.
However, the values are consistent with a porous structure for the fullerene clusters, which would reduce the effective spin density in the cluster.
The structure of fullerene clusters depends on the formation process.\cite{Makhmanov2016}
Slow aggregation leads to densely packed clusters, whereas mechanical agitation\cite{Makhmanov2016} and exposure to light\cite{Dattani2015} lead to the formation of fractal clusters.
It is therefore possible that the concentration procedure, during which nitrogen was bubbled through the solution under ambient light conditions, promoted the formation of porous clusters.
Furthermore, the \fnatcs~sample used in this work was stored under ambient conditions for approximately one year after the initial purification procedure.
The fullerene molecules may have oxidised during this storage period, which further promotes the formation of large clusters by forming epoxide bonds between individual fullerenes.\cite{Dattani2015}

This rotating cluster model explains why the observed linewidth is much greater than the minimum achievable linewidth set by relaxation mechanisms inherent to the molecule itself, such as the Orbach process.\cite{Morton2006}
The narrowest linewidths measured previously required careful sample preparation to inhibit clustering, which occurs at concentrations above 0.06\,mg/mL,\cite{Morton2007} and minimise paramagnetic impurities such as dissolved oxygen.\cite{Morton2007}
However, using such a low concentration in our experiments was not feasible given the small sample volume of the pressure cell and the available sample purity.

The field-independent clock transition in the low-field spectrum of \fnatcs~should suppress the dipolar broadening.\cite{Wolfowicz2013,Shiddiq2016}
However, the clock transition does not protect against all dipolar decoherence mechanisms.\cite{Wolfowicz2013}
The maximum achievable stability of a fullerene clock could therefore be constrained by the need to compromise between increasing spin density to increase signal intensity and the need to minimise linewidth broadening caused by dipole--dipole interactions.\cite{PhysRevLett.119.140801}

\section{Conclusions}

The proposed model, based on van der Waals interactions between the nitrogen atom and the cage, explains the pressure dependence of the \natcs~hyperfine coupling.
The model predicts a smaller pressure shift than is observed experimentally, which may indicate that considering only van der Waals interactions between the nitrogen atom and the cage is insufficient.
However, the model provides reasonable agreement with the data given the spread in predictions of the bulk modulus.
While the small magnitude of the pressure shift likely precludes using it to offset the temperature shift of the clock frequency, it ensures that a \fnatcs~based frequency reference is insensitive to atmospheric pressure fluctuations.

The pressure-dependent linewidth is well fitted by a model based on dipole--dipole interactions between \fnatcs~spins embedded in fullerene clusters.
Dipolar broadening of the spin resonance depends on pressure via the viscosity of the solvent, which modifies the rotational correlation time of the cluster.
This model provides insight into spin relaxation processes in concentrated \natcs~solutions that may limit the stability of a fullerene clock.
\section*{Acknowledgements}
We thank Professor D.~M.~Murphy for the use of the high-pressure ESR spectrometer, which was supported by EPSRC (EP/K017322). We also thank Dr.~W.~M.~Myers for helpful discussions.
We acknowledge DSTL, EPSRC (EP/J015067/1, EP/K030108/1, EP/N014995/1), and the Royal Academy of Engineering.
\end{document}